\documentclass[a4paper, 11pt, notitlepage]{scrartcl}
\usepackage[margin=1in]{geometry}                
\usepackage[parfill]{parskip}    

\usepackage[T1]{fontenc}
\usepackage{libertine}

\usepackage{amsmath}
\usepackage{amssymb}
\usepackage{graphicx}
\usepackage{setspace}
\usepackage{xcolor}
\usepackage{authblk}

\usepackage{cite}

\newcommand{\norm}[1]{||#1||}

\newcommand{\be}{\begin{equation}}

\newcommand{\ee}{\end{equation}}
\newcommand{\bes}{\begin{eqnarray}}
\newcommand{\ees}{\end{eqnarray}}

\newcommand{\figref}[1]{Figure~\ref{#1}}
\renewcommand{\eqref}[1]{Eq.~(\ref{#1})}

\newcommand{\avg}[1]{\langle #1 \rangle}

\title{Quantifying the impact of network structure on speed and accuracy in collective decision-making}

\date{}                                           

\author[1]{Bryan C.~Daniels}
\author[2,3]{Pawel Romanczuk}
\affil[1]{ASU--SFI Center for Biosocial Complex Systems, Arizona State University, Tempe, Arizona, USA}
\affil[2]{Institute for Theoretical Biology, Department of Biology, Humboldt Universit\"at zu Berlin, Germany}
\affil[3]{Bernstein Center for Computational Neuroscience, Berlin, Germany}

\begin{document}

\maketitle

\begin{abstract}

\noindent 



Found in varied contexts from neurons to ants to fish, 
binary decision-making is
one of the simplest forms of collective computation.
In this process, information collected by individuals
about an uncertain environment is accumulated to 
guide behavior at the aggregate scale.
We study binary decision-making dynamics in networks
responding to inputs with
small signal-to-noise ratios, looking for quantitative
measures of collectivity that control decision-making
performance.
We find that decision accuracy 
is controlled largely by three factors: the leading eigenvalue of the
network adjacency matrix, the corresponding eigenvector's
participation ratio, and distance from the corresponding symmetry-breaking
bifurcation.  
This allows us to predict how decision-making
performance scales in large networks based on their spectral properties.
Specifically, we explore the effects of 
localization caused by the hierarchical assortative
structure of a ``rich club'' topology.
This gives insight into the 
tradeoffs involved in the higher-order structure 
found in living networks performing collective
computations.  


\end{abstract}

\section*{Introduction}

Collective intelligence refers to the ability of groups of individual components to process environmental information and successfully perform adaptive functions at a larger collective scale.  Building a coherent framework for understanding distributed functionality is challenging in that the internal structure of natural and engineered collectives varies strongly, from quasi-homogeneous systems like swarms of identical robots, to fish-schools consisting of similarly behaving individuals but with persistent behavioral differences  (``personalities'') \cite{jolles2017consistent,bierbach2018using} or different prior information \cite{couzin2011uninformed,pinkoviezky2018collective}, to strongly heterogeneous and hierarchical systems like primate societies \cite{daniels2012sparse,daniels2017control} or neurons in a brain \cite{HarVanSpo12,NigShiIto16}.   Facing this diversity, a key challenge for building a better abstract understanding of collective intelligence is to determine which details of such systems are most important to collective function and which are incidental and can be ignored. In this way, we are searching for measures that usefully quantify ``collectivity'' across a broad continuum of complex systems.

In addition to diversity in heterogeneity and communication structure, myriad types of functions may be implemented in a collective system, ranging in complexity from simple majority consensus to high-level abstract information processing.  
Here we focus on a particularly simple function---making a correct binary decision about the sign of a noisy distributed input---and look for network statistics that delineate the full range of strategies that can be used to successfully perform this collective function.

Past experimental investigations of collective decision-making have mostly not 
addressed network structure, instead assuming all-to-all coupling 
and focusing on optimal rules for aggregating decisions made by individuals \cite{conradt2005consensus,dyer2008leadership,wolf2013accurate,ColFlaSer06,juni2015flexible}. However, an increasing number of studies are beginning to investigate non-trivial network structures \cite{KeaJudTan09,judd2010behavioral,rosenthal2015revealing,navajas2018aggregated}. For example, Kearns et al~\cite{KeaJudTan09} look at the effect of varying network structure in consensus formation in human groups.  They find, e.g., that ``preferential attachment'' networks lead to faster consensus than Erdos--Renyi. 

Theoretically, many examples of collective decision-making can be effectively described using networks of coupled dynamical components.  Structural properties of such networks and how they affect self-organization and collective behavior have long been a focus of complex systems research (see e.g. \cite{olfati2004consensus,arenas2008synchronization,gross2007adaptive,kozma2008consensus}). Particularly well studied are effects of network structure on emergent dynamics in the context of synchronization and consensus formation \cite{olfati2004consensus,arenas2008synchronization}. 
Theoretical research often aims to map the phase diagram of system dynamics as a function of underlying network structure parameters, for example to identify regions of synchronized versus random dynamics (see e.g. \cite{arenas2008synchronization} and references therein). This language of phase diagrams, originating in statistical physics, has also been used to hypothesize that in order to ensure optimal information processing, collective systems should operate near phase transitions (critical manifolds) \cite{mora2011biological,carbone2015model}.  \footnote{In some large $N$ limits, hierarchical modular networks can have infinitely many localized modes corresponding to critical coupling strength values over a continuous range---this produces a so-called ``Griffiths phase'' \cite{MorMun13}.  We do not focus on this here because we anticipate our methods will be most useful applied to known finite networks.}

Corresponding theoretical insights have driven the systematic analysis of structural properties of artificial and real-world networks, including node and degree heterogeneity \cite{kumar2010structure} and structural hierarchies \cite{girvan2002community,boccaletti2007detecting}.
In particular, many real-world collective systems exhibit 
a ``rich club'' (core-periphery) structure, 
with examples coming from neuroscience \cite{VanSpo11,HarVanSpo12,BasWymRom13,GolZalHut15,NigShiIto16}, 
social science, and biochemistry \cite{ColFlaSer06,LeoFleAlv16}. 
The rich club refers to a subset of nodes that a) have a larger (in-)degree and b) are more likely to be connected to other rich club nodes than in an otherwise random wiring. It has been argued that such a core-periphery topology may play an important role for the function of complex information processing systems (see e.g. \cite{HarVanSpo12,senden2014rich,GolZalHut15}). 

The dynamical effects of network structure have been explored largely in the 
context of synchronization or consensus, the problem of collective agreement.
Extending to the problem of decision-making  
also requires a notion of correctness: we want a system that
not only produces collective agreement on any consensus state, but on the correct state, given a source of input information.  Binary collective decision-making in this sense 
maps naturally onto ``noisy integrator'' models, such as leaky integration to bound
(Ornstein--Ullenbeck \cite{srivastava2014collective}) and related models with stable attractors representing decision states \cite{Wan02,DanFlaKra17}.
A general constraint for any such decision-making system is the tradeoff between speed and accuracy \cite{ChiSkoRai09,MarBogDor09,srivastava2014collective}. Recently, it has been shown that the speed--accuracy tradeoff in simple collective decision models can be quantified in terms of distance from a bifurcation \cite{DanFlaKra17,AreJinDan18}. 

Motivated by the above findings, we 
focus in this work on the question of how a rich-club structure affects 
the speed and accuracy of collective decision-making.
In particular, we look for network statistics that capture the most important properties controlling collective performance in decision dynamics.

\section*{Results}

\begin{figure}
\centering
\includegraphics[scale=0.70]{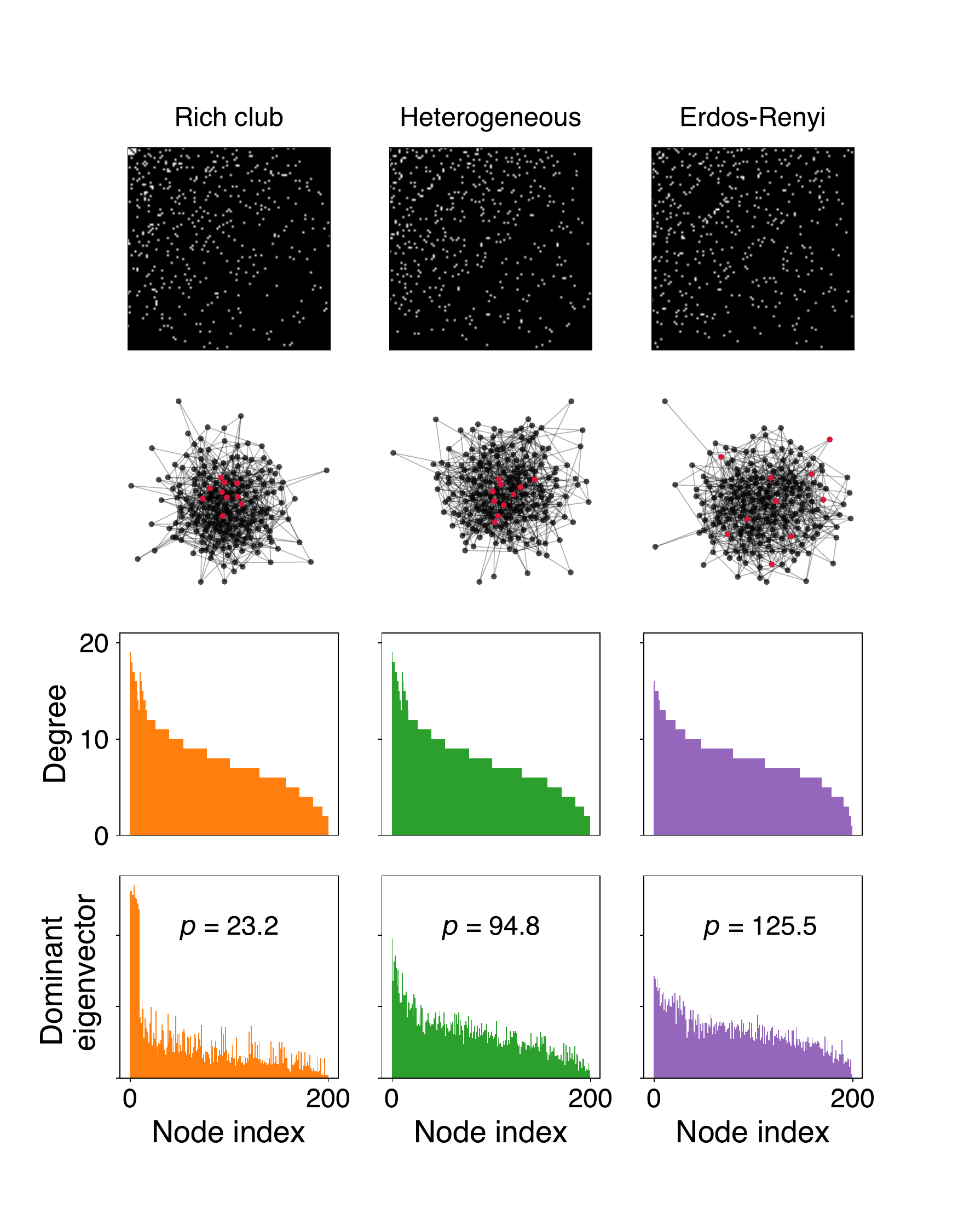}
\caption{ \textbf{Three random networks with varying higher-order network structure.} 
All networks share the same size ($N = 200$) and total number of edges (800).
The ``rich club'' network is generated such that a core group of 
10 nodes have increased probability
of edges within the group ($b_\mathrm{rich} = 257$); 
the ``heterogeneous'' network has identical degree distribution
to the rich club network but with edges otherwise randomized; and the ``Erdos--Renyi''
network is selected such that all possible edges have equal probability.  
We characterize each network using its adjacency matrix (top row), network diagram
(second row), degree distribution (third row), and dominant eigenvector and
corresponding participation ratio $p$ (bottom row).
Throughout the figure, nodes are ordered first by core and periphery groups and then sorted by degree.   Core nodes are highlighted in red in the network diagrams.
\label{networksFig}%
}
\end{figure}

\subsection*{Collective decision-making model}

A simple minimal model of distributed  
decision-making defines dynamics for
the internal noisy states of individual components, 
each of which
receives the same input signal $I$, 
recovers to a null state on a timescale $\tau$, 
and is affected by its neighbors through a 
saturating function of its neighbors' states \cite{DanFlaKra17,AreJinDan18}:
\begin{equation}
\label{dynamicsEqn}
\frac{d s_i}{dt} = - \frac{s_i}{\tau} + \frac{\mu}{\tau} \sum_j A_{ij} \tanh(s_j) + \frac{I}{\tau} + \xi,
\end{equation}
where $I$ is an input signal given uniformly to every node
and $\xi$ is uncorrelated Gaussian noise with
$\langle \xi(t) \xi(t+\Delta t) \rangle = \sigma^2 \tau^{-1} \delta(\Delta t)$.
We explicitly write the differential equations in terms of 
an overall timescale $\tau$; 
in describing neural dynamics, for example,
we expect $\tau$ to be on the order of tens of milliseconds.

We initialize the system in a state $\vec s_0$ that,
in the case of zero noise,
corresponds to
a fixed point undergoing a pitchfork bifurcation
as a function of the coupling strength $\mu$.
This bifurcation separates the case of a 
single stable fixed point at $\vec s_0$ 
and the case of two distinct stable fixed points at 
$\vec s_0 \pm \epsilon \hat e_c$, which we treat as 
decision states (where $\hat e_c$ is the unit 
vector pointing in the direction in which the
decision states emerge from $\vec s_0$ at the
bifurcation).
We focus here on the simplest such bifurcation,\footnote{
    Tuning a second control parameter can locate
    more general pitchfork bifurcations; 
    see \cite{AreJinDan18}.
} 
which occurs at $\vec s_0 = \vec 0$.

To test how the existence of higher-order structure changes the
decision-making performance, we vary the adjacency 
matrix $A$ to
test symmetric networks with fixed size $N$ 
and total number of edges, changing only
the degree distribution and higher-order structure (\figref{networksFig}). 
First, a ``rich club'' network is created through a random generation of a fixed number of edges from
the $N(N-1)$ possible edges, where edges between $N_\mathrm{rich}$ core nodes are biased to be more likely to appear by a factor $b_\mathrm{rich}$.
A corresponding network that has exactly the same degree distribution 
but no rich club is then created by randomly swapping existing edges.
Finally, we test an Erdos--Renyi variant in which all possible edges
are equally likely.
We use the same three specific example networks shown
in \figref{networksFig} in simulations throughout the
paper; results are qualitatively similar for other 
networks sampled from each ensemble.

\subsection*{Speed--accuracy tradeoff}

\begin{figure}
\centering
\includegraphics[scale=0.6]{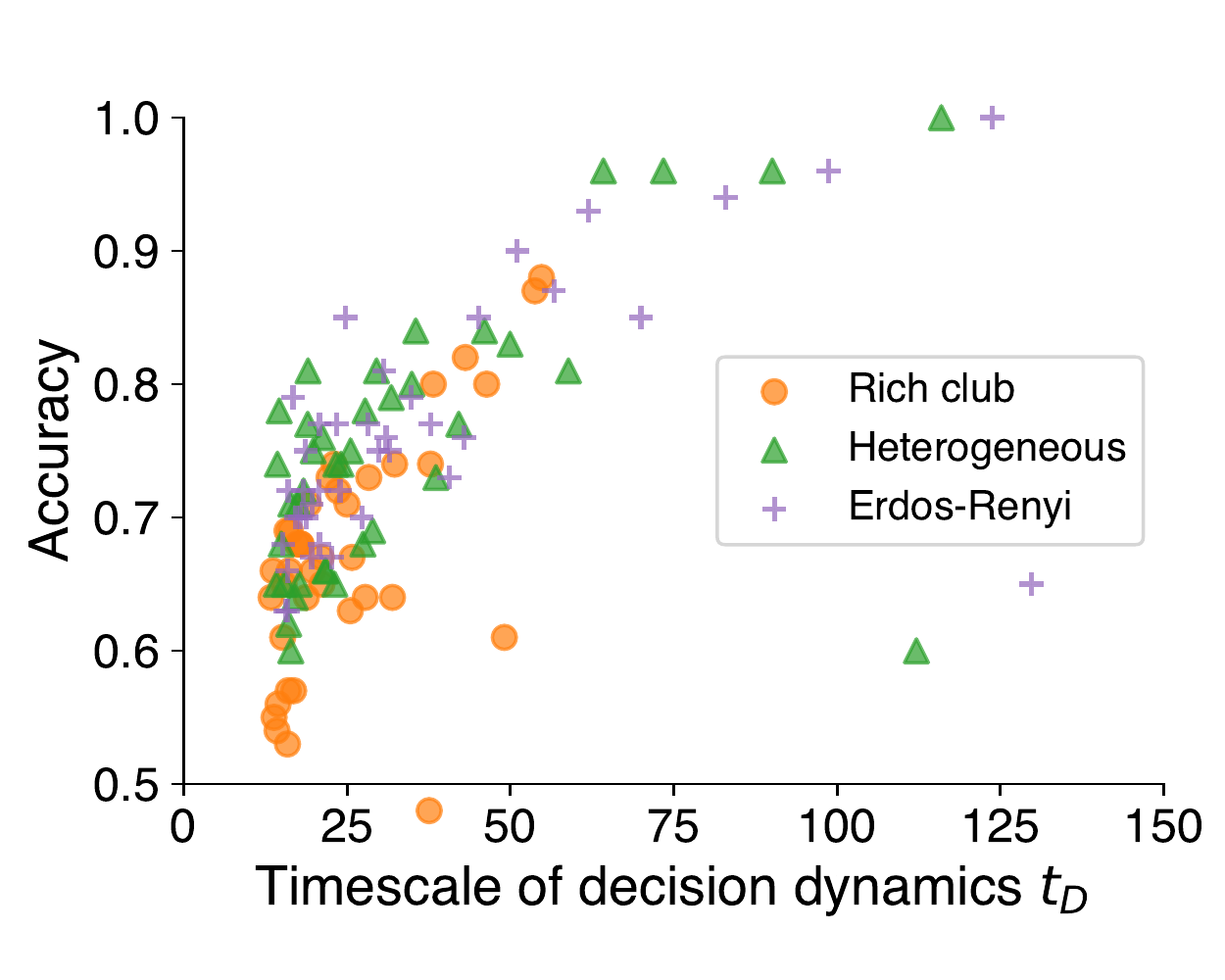}
\caption{ \textbf{Speed--accuracy tradeoff.}
Simulations show that maximal accuracy of a noisy collective decision occurs when the decision process happens over a longer timescale, lengthening the time over which the system remains sensitive to the input.  Here and in other plots unless otherwise specified, $\tau = 1$, $\sigma = 0.05$, $I = 0.001$, the duration of the signal $= 200 \tau$, the total simulation time $= 2000 \tau$, $\Delta \mu$ ranges from $0$ to $0.04$, and we average over 100  simulations for each network. 
\label{speedAccuracyFig}%
}
\end{figure}

We test each network in its ability to integrate information about a
signal that is small compared to the noise and then retain that information
after the signal is removed.
We define accuracy in terms of whether the system
ends the simulation near the correct decision state,
the one that 
lies in the direction of the input $+I$ (instead of $-I$).
Specifically, we test whether the sign of the final state  along the unstable dimension, $(\vec s_\mathrm{final} - \vec s_0) \cdot \hat e_c$, is the same as the sign of the input along that dimension, $I \vec 1 \cdot \hat e_c$.

As in previous studies using a similar model \cite{DanFlaKra17,AreJinDan18}
(and across a wide variety of systems in general \cite{FraDorFit03,ChiSkoRai09,MarBogDor09}),
we expect to find a speed--accuracy tradeoff:
Slow dynamics should produce better accuracy, as the system is able to integrate the input over a longer time before fixating within a single decision state, whereas
fast dynamics produce a decision based primarily on noise.

To quantify the speed of the decision, 
we measure a characteristic time $t_D$
over which the system approaches the final decision
fixed point.
We first define the two decision states $\vec s_{\pm}^*$ as 
the stable fixed points of the dynamics in 
\eqref{dynamicsEqn} with zero input and zero noise.\footnote{
    In the case here with $\vec s_0 = \vec 0$, the two decision states are
    related simply by an inversion symmetry:
    $\vec s_+^* = -\vec s_-^*$.  Sufficiently close
    to the bifurcation, we expect analogous results
    for the speed--accuracy tradeoff even in more 
    complicated cases where this symmetry does not hold \cite{AreJinDan18}.}
We then define the decision timescale $t_D$ as the first time the state $\vec s$ reaches halfway to the decision state $\vec s^*$ along the dimension $\hat s^* = (\vec s^* - \vec s_0)/|\vec s^* - \vec s_0|$.

As expected, our simulations show 
a speed--accuracy tradeoff as we vary 
the overall connection strength $\mu$, 
shown in \figref{speedAccuracyFig}. 
When tuned to a given decision timescale, the
accuracy is largely unaffected by network structure.
The highest accuracy is observed when the system supports long timescale dynamics.

Given that performance is largely controlled by the decision timescale $t_D$, we would like to understand how the network structure, defined by the adjacency matrix $A$, controls $t_D$.  We expect that the 
largest timescales will occur near the symmetry-breaking
transition that creates the two decision states.

\subsection*{Locating the transition and decision states}

First, we must locate the relevant pitchfork bifurcation,
which controls the transition between
dynamics in which node states are not correlated over long times 
(when $\mu$ is small and interactions between nodes are weak) into
dynamics in which a nodes can collectively store a long-term memory
(when $\mu$ is large enough that 
interactions support a self-reinforcing consensus state).
With zero input and zero noise, it is straightforward to find this 
transition, 
by analyzing how a small perturbation 
$\delta \vec{s}$ to the initial state $\vec s_0$ changes under the dynamics:
\begin{equation}
\tau \frac{d \delta \vec{s}}{dt} = - \delta\vec{s} + \mu A\delta\vec{s} =(\mu A - \mathcal{I})\delta\vec{s},
\end{equation}
where $\mathcal{I}$ is the identity matrix.  Then the behavior is most easily analyzed in
the basis of eigenvectors of $A$: the dynamics 
along eigenvector $\hat e_\lambda$ are given by
\begin{equation}
\tau \frac{d \delta \hat e_\lambda}{dt} = (\mu \lambda - 1) \delta \hat e_\lambda.
\end{equation}
Thus the initial state $\vec s_0$ will be stable until $\mu$ becomes large enough to make
$(\mu \lambda - 1)$ positive.  In other words, as we expect from basic
linear stability analysis \cite{New10}, the critical value of $\mu$ at which
$\vec s_0$ first becomes unstable is controlled by 
the largest eigenvalue $\lambda_c$ of $A$:
\begin{equation}
\mu_c = \lambda_c^{-1}.
\end{equation}
The symmetry between positive and negative values of $s$ means
that this is a pitchfork bifurcation, and two stable fixed points
emerge from the unstable fixed point along the dimension of the
eigenvector corresponding to $\lambda_c$.

The distance between each stable fixed point
(decision state) and the unstable starting point $\vec s_0$
grows as a function of $\mu$:
Near the transition [see Appendix \eqref{nustarEqn}],
\begin{equation}
\label{localApprox}
\norm{\vec s^* - \vec s_0} \approx 
	\sqrt{ 3 \Delta \mu \lambda_c p}
    = \sqrt{ 3 \bar \mu p},
\end{equation}
where $\Delta \mu = \mu - \mu_c$, $\bar \mu = \Delta \mu / \mu_c$ is the reduced 
distance from the transition, and $p = 1/|(\hat e_c)^4|$ 
characterizes the distributedness of the
eigenvector $\hat e_c$ corresponding to the leading
eigenvalue $\lambda_c$.

\begin{figure}
\centering
\includegraphics[scale=0.55]{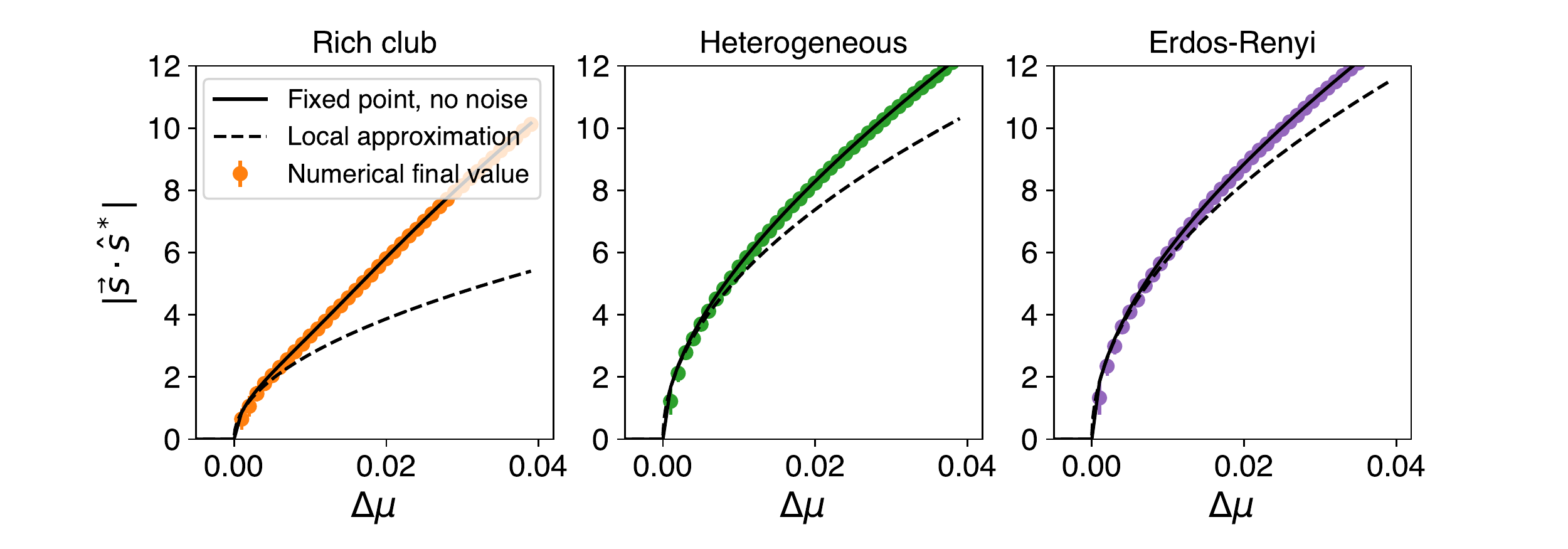}
\caption{ \textbf{A local approximation for the location of decision state fixed points.}
Analytical approximations are 
compared to the simulated norm of the 
steady-state vector versus coupling strength.
A local approximation using only the eigenvalue $\lambda_c$ and participation ratio $p$ [\eqref{localApprox}] is shown as a dashed line.
The numerical solution of the zero input, zero noise case [\eqref{dynamicsEqn} with $I = \sigma^2 = 0$] is shown as a solid line.  The average final state in simulations, including noise but no input, is shown as colored points, with errorbars (not visible for most points) corresponding to standard deviation of the mean.  
\label{fixedPointDistanceFig}%
}
\end{figure}

The value $p$ sets
the scale of the distance between the 
collective decision states, and it
corresponds roughly to the number of 
individual nodes contributing to the mode (see the
bottom row of \figref{networksFig}).
In general, $p$ varies between 1 for a completely
localized mode and $N$ for a 
completely delocalized mode (produced, for example,
by homogeneous all-to-all coupling).
The inverse of $p$ appears in studies of localization in
random matrix theory, where it has been called 
the ``inverse participation ratio'' \cite{PleGopRos02,MetNerBol10};
we therefore call $p$ the participation ratio.

\figref{fixedPointDistanceFig} compares this zero-noise local approximation
to the zero-noise numerical solution for the fixed point $\vec s^*$ and to
the final state of the simulation including noise.\footnote{
    Noise also affects the location of the transition.
    In our simulations here we use a small noise parameter
    $\sigma$ (with input signal $I$ even smaller to 
    achieve a small signal-to-noise ratio); this means
    the effect of noise on the transition location is 
    minimal on the scales we test.  We calculate the
    lowest-order correction to $\mu_c$ in the Appendix
    and find that it is on the order of $10^{-3}$ for
    the plotted cases.
}  For each network,
the transition occurs at the expected $\mu_c$ and with the expected local
dependence on $\Delta \mu$.  Because the largest timescales also occur
near the transition, this local analysis will allow us to 
approximate the maximal decision timescale in the next section.\footnote{
    Note that, in the rich club network, increasing the coupling beyond the
    scale shown in \figref{fixedPointDistanceFig} can also create bistability in the peripheral nodes.  These cases have four stable fixed points,
    two of which correspond to the core and periphery nodes coming to consensus on
    conflicting decisions, and two in which core and periphery disagree. 
    In our current setup, these cases do not change our analysis because 
    the core always decides first, biasing the remainder of the system.  
}

\subsection*{Predicting the timescale of the decision}

\begin{figure}
\centering
\includegraphics[scale=0.55]{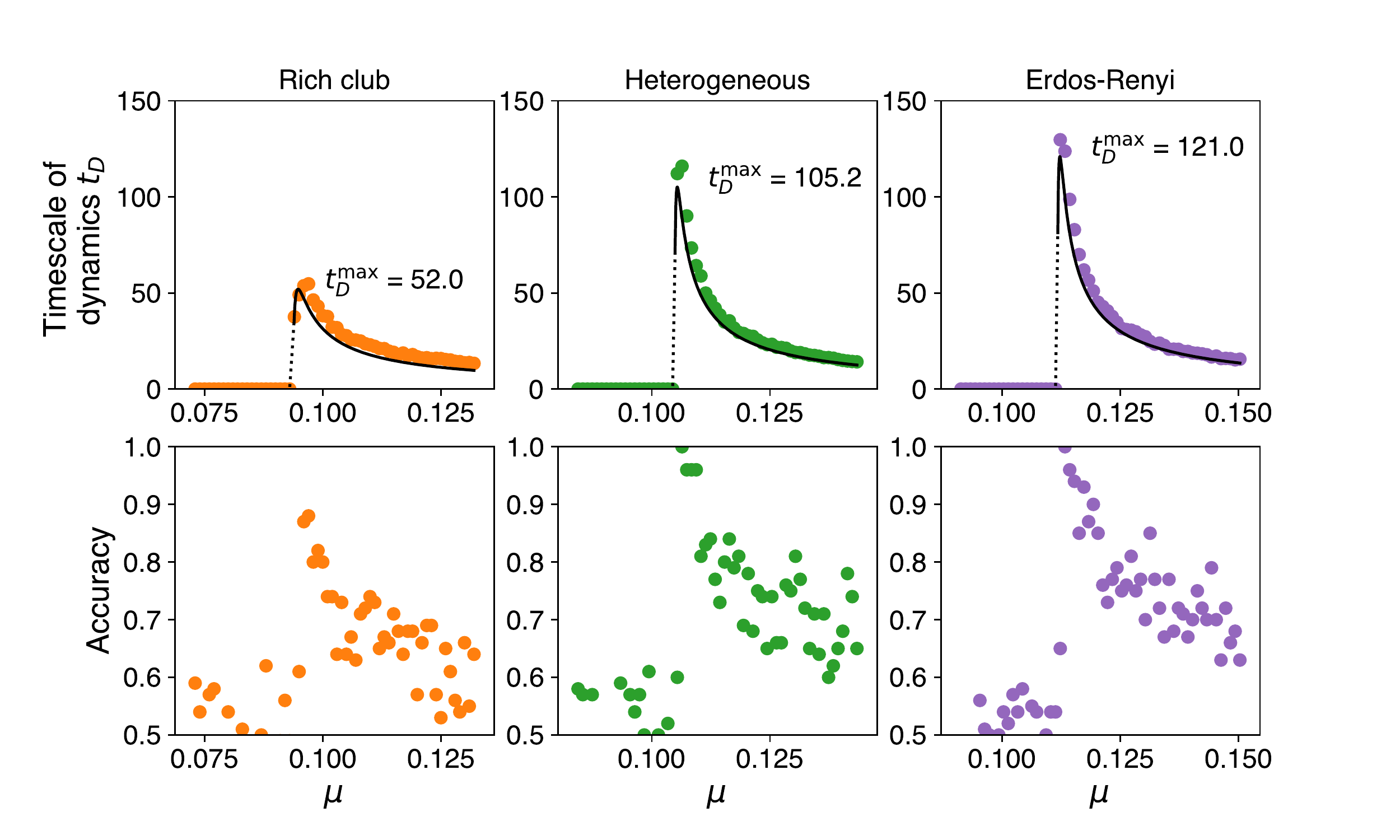}
\caption{ \textbf{Predicting the timescale of the decision.}  
Critical slowing down increases the timescale of motion
toward the decision state
(top row) when the coupling strength $\mu$ is tuned near the 
symmetry-breaking bifurcation.  Simulation results are shown
as colored points, and an analytical approximation of $t_D$ 
[\eqref{timescaleEqn}] is shown
as dashed and solid black lines.
The approximate maximum timescale $t_D^\mathrm{max}$,
expected to set an upper limit on accuracy, is
proportional to $\sqrt{p}$.
Accuracy also peaks near the transition (bottom row).
\label{timescaleFig}%
}
\end{figure}

In the absence of noise, 
the timescale of the decision is expected to
diverge at the transition, because the 
fixed point at the origin becomes marginally
stable.    With noise, this is smoothed
out in a predictable way, leading to a simple
equation for the timescale derived in the 
Appendix, \eqref{timescaleEqn}.
Roughly, the timescale is determined by a 
combination of the distance between
the two decision states (proportional to $\sqrt{\Delta \mu}$), 
the characteristic timescale of exponential
growth away from the unstable fixed point
(proportional to $\Delta \mu^{-1}$), and the
characteristic speed of motion due solely to
noise (proportional to $\sigma$).

In \figref{timescaleFig}, we demonstrate that
 the decision timescale is well-approximated by
 \eqref{timescaleEqn}.  As we saw before in 
 \figref{speedAccuracyFig}, longer timescales
correspond to better accuracy.  
Further, this analysis allows us to predict the
maximal decision timescale supported by a given
network under a given level of noise $\sigma$; we
find [see Appendix \eqref{tDmaxEqn}]
\begin{equation}
t^\mathrm{max}_D \propto \sqrt{p} / \sigma;
\end{equation}
that is, the timescale supported by the collective mode
scales with the square root of the 
participation ratio.  Consequently, due to the fundamental speed--accuracy tradeoff, $p$ becomes a useful quantification of higher order network structure that sets a limit on collective decision accuracy.




\section*{Discussion}

We study here a collective decision process that relies on the phenomenon of critical slowing down, a mechanism for creating long-timescale dynamics. 
In the case of small signal-to-noise ratio,  
decision accuracy is limited by the 
timescale of collective dynamics, and the system
must be tuned near a symmetry-breaking bifurcation to
successfully integrate information into an accurate decision.  
Varying the distance from the bifurcation $\Delta \mu$ 
traces out a speed--accuracy tradeoff (\figref{speedAccuracyFig}).

In the spirit of quantifying collectivity, our aim is to
characterize the aspects of network connectivity that 
control this timescale and therefore place limits on 
collective decision accuracy.
We find that the most important factors characterize the normal mode of the network that is least stable.  This is the mode that first 
becomes unstable as interaction strengths are increased, 
thereby leading to bistability that encodes a binary decision.
First, the leading eigenvalue $\lambda_c$, effectively a measure of the connectivity of individuals participating in the mode, sets the scale of the critical coupling $\mu_c$ required for reinforcing the decision state.  Second, the participation ratio $p$ of the corresponding eigenvector is a measure of the number of individuals participating in the mode.

Our main result is to identify $\lambda_c$ and $p$ as important measures 
for quantifying collective behavior in heterogeneous networks.
Given any detailed network structure, 
these two simple statistics encapsulate 
the network's ability to create long-timescale dynamics.
This allows us to predict how 
collective timescales behave across a variety of network
structures.
For instance, the maximal decision timescale that can be produced by the critical slowing mechanism increases in a predictable 
way as more individual nodes are allowed to participate in the 
unstable mode, scaling as $\sqrt{p}$.  This fits with our rough intuition, as we expect that the effects of noise will shrink in a group of $N$ individuals as $1/\sqrt{N}$.

Generally, our results resonate with recent studies 
that focus on low-dimensional collective modes controlling the most
important aspects of distributed computation in biological networks \cite{GanBisRoi08,MasOst18}.
The importance of the principal eigenvalue and corresponding (inverse) participation ratio 
hints at connections between our model of decision-making and 
related characterizations of disease spreading \cite{GolDorOli12}, 
correlations in financial data \cite{PleGopRos02}, and Anderson localization in condensed matter physics \cite{FyoMir92}.

Our motivation began with understanding the functional 
consequences of rich-club structure and criticality in 
the brain.  These results allow us
to speculate about fundamental tradeoffs:
What are potential advantages and disadvantages to hierarchical rich-club structure?  On 
the one hand, more distributed connectivity may be advantageous in that it leads to more distributed collective modes, longer timescales,
and therefore better averages over the noisy knowledge of
individuals.  
On the other hand, the localized modes created by a rich club structure could be advantageous for modularized function 
and localized control.  In this way, the rich club
could be a way to bring only a subset of the system supercritical,
with consequently reduced noise-reduction benefits of collectivity.


We expect this framework and intuition to be useful in systems in which
the interaction structure remains fixed over the timescale of 
a single decision process, but may vary over longer adaptive
timescales.  Besides neural dynamics, such a framework may be useful 
for describing genetic regulatory networks producing cell fate decisions 
during development \cite{MojSkuZho16},
social networks producing consensus about dominance hierarchies \cite{BruKraFla13},
and networks of influence underlying decisions by political bodies \cite{LeeBroBia15,BarHuaSpa18}.
In such systems, the computation of decisions happens on relatively fast timescales, while on longer adaptive timescales,
there may be tuning of the network that could change the relevant parameters 
$\lambda_c$ and $p$.

To guide our intuition, our analysis focused on the simplest symmetry-breaking
bifurcation, where the initial state of each individual is the same 
($\vec s_0 = \vec 0$).  It will be useful in future work to focus on more
complicated transitions (as explored in \cite{AreJinDan18}), where 
we expect that differing states and therefore saturations across individuals 
will modify the calculation, perhaps leading to a generalized form of the
participation ratio that weights individuals by their contributions.


%

\section*{Acknowledgments}

PR acknowledges funding by the Deutsche Forschungsgemeinschaft (DFG, German Research Foundation) under Germany's Excellence Strategy -- EXC 2002/1 "Science of Intelligence" -- project number 390523135, as well as through the Emmy Noether program, project number RO4766/2-1.

\section*{Appendix}

\subsection*{Derivation of distance between stable fixed points}

The normal form of a system undergoing a pitchfork bifurcation is 
\begin{equation}
\label{expansion}
\frac{d\nu}{dt} = a \nu + b \frac{\nu^3}{6}.
\end{equation}
In a one-dimensional system with state $x$ and dynamics $d x / dt = F(x)$ that has a pitchfork bifurcation at $x = x_0$, 
the system is described by \eqref{expansion} near $x_0$, with
$\nu = x - x_0$, $a = dF(x)/dx |_{x = x_0}$, and $b = d^3 F(x)/dx^3 |_{x = x_0}$.
This is the Taylor series of $F(x)$ at $x = x_0$ up to third order,
where the second-order term disappears due to the symmetry 
that is required for a pitchfork bifurcation: 
$F(x_0 + \delta) = -F(x_0 - \delta)$ near $\delta=0$.
The bifurcation happens when $a$ changes sign.  We focus here on the case that creates two stable fixed points (decision states), which coincides with $b < 0$.

Solving \eqref{expansion} for $d\nu/dt = 0$, we find one fixed 
point at $\nu^* = \nu_0$ that changes from stable 
when $a < 0$ to unstable when $a > 0$, and two 
stable fixed points when $a > 0$ at 
\begin{equation}
\label{nustarEqn}
\nu^* = \pm \sqrt{ 6a / |b| }.
\end{equation}
For example, in a simple one-dimensional case where 
$F(x) = -x + \mu \tanh{x}$, we have $\nu = x$, $x_0 = 0$, $a=\mu-1$, and
$b=-2\mu$.  Inserting into \eqref{nustarEqn},
we find, for small $\Delta \mu \equiv \mu - 1$, 
$\nu^* \approx \pm \sqrt{3 \Delta \mu}$.

In the higher-dimensional context of \eqref{dynamicsEqn},
$\nu$ becomes the linear combination of state $\vec s$ along the
dimension of the least-stable dimension $\hat e_c$:
$\nu = \vec s \cdot \hat e_c$.
Then, to produce the Taylor series corresponding to 
\eqref{expansion}, we take the relevant directional 
derivatives of the right-hand side of \eqref{dynamicsEqn}.
Calling the zero-noise, zero-input part of the dynamics
$\vec F$ [that is, $F_i(\vec s) = - s_i + \mu \sum_j A_{ij} \tanh(s_j)$],
we have
\begin{eqnarray}
a \,\,\, =& \hat e_c \cdot \frac{\partial \vec F(\vec s)}{\partial \nu}\big|_{\vec s=\vec 0} 
  &= \hat e_c \cdot (\nabla F \cdot \hat e_c)|_{\vec s=\vec 0} = \mu \, \hat e_c ^T A \hat e_c - 1 = \mu \lambda_c - 1; \\
b \,\,\, =& \hat e_c \cdot \frac{\partial^3 \vec F(\vec s)}{\partial \nu^3}\big|_{\vec s=\vec 0} 
  &= \hat e_c \cdot ( \nabla ( \nabla ( \nabla F \cdot \hat e_c ) \cdot \hat e_c ) \cdot \hat e_c ) |_{\vec s=\vec 0} = -2 \mu \lambda \sum_i (\hat e_c)_i^4.
\end{eqnarray}
Inserting this into \eqref{nustarEqn} produces
\begin{equation}
\nu^* = \pm \sqrt{3p\frac{\mu - \mu_c}{\mu}} = \pm \sqrt{3\bar \mu p} + O(\Delta \mu^{3/2}), 
\end{equation}
where $\mu_c = 1/\lambda_c$, $p = 1/\sum_i (\hat e_c)_i^4$, $\Delta \mu = \mu - \mu_c$, and $\bar \mu = \Delta \mu / \mu_c$.

We note that the above result assumes that the adjacency matrix $A$ is symmetric.  In the asymmetric case, 
\begin{equation}
\nu^* \approx \sqrt{ 3 \bar \mu \frac{\lambda_c}{(A^T \cdot \hat e_c) \cdot (\hat e_c)^3 } },
\end{equation}
where $\hat e_c$ is the normalized eigenvector of $A$
corresponding to $\lambda_c$.


\subsection*{Derivation of approximate decision timescale $t_D$}

We define the decision timescale $t_D$ as
the time for $\nu = |\vec s \cdot \hat s^*|$ 
to reach halfway to the fixed point $\nu^*$.
To approximate $t_D$
in the presence of noise, we patch together two
types of behavior.  First, for sufficiently 
small times $t$, we expect the average behavior
along $\hat e_c$ to be dominated by noise [the $\xi$ term
dominates in \eqref{dynamicsEqn}].
Noise dominates here
because the system is still close to the
fixed point at the origin, where, at the bifurcation,
the first two terms cancel up to second order in $\nu$.  
Neglecting all terms other than the noise term,
the average behavior is given by 
$\avg{\nu_1(t)} = \sigma \sqrt{t/\tau}$.
Then, after a crossing time $t_\mathrm{cross}$,
the first two terms dominate 
and noise becomes unimportant.
Now neglecting the noise term, given an
initial condition of 
$\nu_0 = \avg{\nu_1(t_\mathrm{cross})}$, and considering
for simplicity only the lowest-order approximation
of $F$ near the unstable fixed point,
the state simply grows exponentially:
$\avg{\nu_2(t)} = \nu_0 \exp (t-t_\mathrm{cross})\bar \mu/\tau$.

We patch these two solutions together by
defining $t_\mathrm{cross}$ as the time when
their derivative matches:
$$d\avg{\nu_1(t)}/dt|_{t = t_\mathrm{cross}} 
 = d\avg{\nu_2(t)}/dt|_{t = t_\mathrm{cross}}\ .$$
Solving this produces 
$t_\mathrm{cross} = y \tau/\bar \mu$, where
$y \approx 0.3517$ is the solution to $2 \exp y = y^{-1}$.
Finally, we solve for $t_D$, the time to reach
$\nu^*/2$, as a function of the reduced distance
from the transition $\bar \mu$:
\begin{eqnarray}
\label{timescaleEqn}
t_D(\bar \mu) = 
\begin{cases}
  \frac{3 p \bar \mu}{4\sigma^2}\tau &\bar \mu < \bar \mu_\mathrm{cross}, \\
   \frac{\tau}{2 \bar \mu} \left( 2y 
      + \log \frac{3 p \bar \mu^2}{4 y \sigma^2} \right) &\bar \mu \geq \bar \mu_\mathrm{cross},
\end{cases}
\end{eqnarray}
where $\bar \mu_\mathrm{cross} = 2 \sigma \sqrt{y/3p}$.
This approximation of the decision timescale is
plotted in \figref{timescaleFig} as dashed and solid lines
(dashed for $\bar \mu < \bar \mu_\mathrm{cross}$ and
solid for $\bar \mu > \bar \mu_\mathrm{cross}$).
The function $t_D(\bar \mu)$ has a maximum
at $\bar \mu_\mathrm{max} = \bar \mu_\mathrm{cross} \exp(1-y)$, producing the maximal decision timescale for a given transition as
\begin{equation}
\label{tDmaxEqn}
t^\mathrm{max}_D = \frac{\tau z \sqrt{3 p}}{2 \sigma}, 
\end{equation}
where $z = e^{y-1}/\sqrt{y} \approx 0.882$.



\subsection*{Impact of noise on the critical point}

In order to asses the general impact of noise on the critical coupling strength $\mu_c$, we consider a mean-field approach (fully connected graph) in the absence of an external signal ($I=0$). The general approach employed here is analogous to the one used in \cite{romanczuk2010collective,grossmann2012active}, where a more detailed account can be found. The mean field stochastic differential equation corresponding to Eq.~\ref{dynamicsEqn} reads:
\begin{align}
\frac{d s_i}{dt} = - \frac{s}{\tau}+\frac{\tilde \mu}{\tau} \langle \tanh s \rangle + \xi
\end{align}
Here $\langle \tanh s \rangle$ represents the expectation value of the interaction term, and $\tilde \mu=N \mu$ is the mean field coupling strength scaled by the number of nodes $N$. Assuming $s\ll 1$, we use the Taylor expansion $\tanh x \approx x-x^3/3$ to rewrite the interaction of the individual node with the mean field in terms of the first and third moment of $s$:
\begin{align}
\frac{d s_i}{dt} = - \frac{s}{\tau}+\frac{\tilde \mu}{\tau} \left(\langle  s \rangle - \frac{1}{3}\langle s^3 \rangle \right)+ \xi
\end{align}
From the above stochastic differential equation we can derive the following nonlinear Fokker-Planck equation for the probability density function (PDF) $p(s,t)=\langle \frac{1}{N}\sum_j \delta(s-s_j) \rangle $:
\begin{align}\label{nlFPE}
\tau \partial_t p(s,t) = -\partial_s\left\{\left[ -s + \tilde\mu \left(\langle  s \rangle - \frac{1}{3}\langle s^3 \rangle \right)\right]p(s,t)\right\}+\frac{\sigma^2}{2}\partial_s^2p(s,t)
\end{align}
Here, the crucial simplifying assumption in the derivation is that the $N$-particle distribution function factorizes, i.e. that the correlations between nodes can be neglected (\emph{mean-field ansatz}).

Inserting Eq. \ref{nlFPE} into
\begin{align}\label{momDefi}
\partial_t \langle s^n \rangle = \partial_t \int_{-\infty}^{\infty} s^n p(s,t) \textnormal{d}s =  \int_{-\infty}^{\infty} s^n \partial_t p(s,t) \textnormal{d}s
\end{align}
produces a hierarchy of coupled evolution equations for the different moments $\langle s^n \rangle$ of the PDF. 

We rewrite the state variable as $s=u+\delta s$, where $u$ is the average state of the system and $\delta s$ is the fluctuation around the mean, and assume that $\langle \delta s^k \rangle=0$ for all odd $k$ ($k=1,3,5,\dots$). 
This allows us to  express the first three moments of $p(s,t)$ as:
\begin{align}\label{moments}
\langle s \rangle & = u \\
\langle s^2 \rangle & = u^2 + \langle \delta s^2 \rangle \\
\langle s^3 \rangle & = u^3 + 3u\langle \delta s^2 \rangle
\end{align}

In the following, we will use the notation  $T=\langle \delta s^2 \rangle$ for the variance of the fluctuations. It can be viewed as an effective ``temperature'' quantifying the intensity of the fluctuations around the mean. 

Eventually combining equations \ref{nlFPE}, \ref{momDefi} and \ref{moments} we arrive at the following evolution equations for the mean $u$ and the temperature $T$:
\begin{align}
    \partial_t u & = u\frac{1}{\tau}\left[\tilde \mu - 1 - \frac{\tilde \mu }{3}\left(u^2+3T\right) \right] \\
    \partial_t T & = \frac{1}{\tau}\left( \sigma^2 -2 T \right)
\end{align}
The corresponding stationary solutions can be easily obtained by solving the above equations for $\partial_t u=\partial_t T =0$. The stationary temperature $T^*$ is 
\begin{align}
    T^*=\frac{\sigma^2}{2}
\end{align}
The cubic equation for the mean yields three stationary solutions $u^*$, which correspond directly to the fixed points $\nu^*$ discussed in the context of the general pitch-fork bifurcation above:
\begin{align}
    u_1^{*} & =0 \\
    u_{2,3}^* & =\pm \sqrt{\frac{3}{\tilde \mu}(\tilde \mu - 1)-\frac{3\sigma^2}{2} }
\end{align}
Here, the $u_1^*$ corresponds to the disordered solution below a critical coupling strength. If the coupling strength becomes too large, then this disordered solution becomes unstable. In the absence of an external signal (bias), we observe a spontaneous symmetry breaking, where $u_{2,3}^*$ correspond to the two possible solutions, identical with the two different branches of the pitchfork. These two stationary solutions exist only if the argument in the square root is positive (as $u^*\in\mathbb{R}$). 

For vanishing noise, $\sigma^2=0$, the critical mean-field coupling strength is $\tilde \mu_c=1$, which is consistent with our previous results if we consider a fully connected graph. For small noise $\sigma\ll 1$ the critical point is modified according to:
\begin{align}
    \tilde \mu_c = \frac{1}{1-\frac{\sigma^2}{2}}\approx 1+\frac{\sigma^2}{2}
\end{align}
Thus, introducing noise leads effectively only to a shift of the critical coupling strength to larger values, without any qualitative change regarding general result obtained for the zero noise case.  Furthermore, the shift is small for the regime we test in this study: with $\sigma = 0.05$ we expect corrections to $\mu_c$ on the order of $10^{-3}$.



\bibliographystyle{unsrt}
\bibliography{rich-club-dynamics}

\end{document}